\documentclass[aip,jmp,amsmath,amssymb,reprint]{revtex4-1}

\usepackage{graphicx}
\usepackage{amsmath,amsfonts,amssymb,bm}
\pdfoutput=1

\begin{document}
	
	\title{Optimized Preprocessing and Machine Learning for Quantitative Raman Spectroscopy in Biology}
	
	\author{E. E. Storey}
	\author{A. S. Helmy}%
	\email{a.helmy@utoronto.ca}
	\affiliation{ 
		Department of Electrical and Computer Engineering, University of Toronto, Toronto, ON M5S 3G4, Canada
	}%
	
	\date{\today}
	
	\begin{abstract}
		Raman spectroscopy's capability to provide meaningful composition predictions is heavily reliant on a pre-processing step to remove insignificant spectral variation. This is crucial in biofluid analysis. Widespread adoption of diagnostics using Raman requires a robust model which can withstand routine spectra discrepancies due to unavoidable variations such as age, diet, and medical background. A wealth of pre-processing methods are available, and it is often up to trial-and-error or user experience to select the method which gives the best results. This process can be incredibly time consuming and inconsistent for multiple operators. 
		
		In this study we detail a method to analyze the statistical variability within a set of training spectra and determine suitability to form a robust model. This allows us to selectively qualify or exclude a pre-processing method, predetermine robustness, and simultaneously identify the number of components which will form the best predictive model. We demonstrate the ability of this technique to improve predictive models of two artificial biological fluids.
		
		Raman spectroscopy is ideal for noninvasive, nondestructive analysis. Routine health monitoring which maximizes comfort is increasingly crucial, particularly in epidemic-level diabetes diagnoses. High variability in spectra of biological samples can hinder Raman's adoption for these methods. Our technique allows the decision of optimal pre-treatment method to be determined for the operator; model performance is no longer a function of user experience. We foresee this statistical technique being an instrumental element to widening the adoption of Raman as a monitoring tool in a field of biofluid analysis.
	\end{abstract}
	
	\maketitle
	
	\section{\label{sec:intro}Introduction}

		An enhancement in existing quantitative analysis capabilities for biofluids using Raman spectroscopy has great potential to improve biomedical diagnostics and monitoring as well as contribute to patient quality of life. Blood, saliva, tears and urine all contain a wealth of vibrational information, which can inform health-care practitioners about a patient's system immunity, waste excretion, nutrient delivery, and many other vital aspects \cite{Kong2015}. Most of these fluids can be collected relatively non-invasively, which contributes to patient comfort during routine or long-term testing. This is in contrast to clinical assessment, which is often the standard technique for analysis and diagnosis, despite its reliance on heavily specialized equipment and trained operators to interpret results. 
	
		There is currently a shortage in the range of diagnostic tools which can flexible to address a range of biofluids, while being simultaneously portable, easy to implement and can be programmed for rapid reliable predictions. Raman spectroscopy is an ideal tool for detailed quantitative analysis of biofluids. It provides unparalleled chemical specificity with minimal sample preparation, and can preserve the fluid under test for further analysis \cite{Enejder2002} \cite{Mak2013}. The challenge associated with Raman approaches is their weak signal, when compared to luminescence. Enhancement techniques such as surface enhanced Raman or Coherent anti-stokes Raman require cumbersome chemical preparation with limited repeatability or expensive, complex instrumentation, respectively. Raman using liquid core waveguides (LCW) can be a powerful technique to enhance the signal from biofluids, with no chemical preparation necessary. Despite the enhancement that attainable using LCW of biofluids, an analytical technique capable of quantitatively handling the widely varied response obtained from biofluids belonging to a range of patients is still lacking.
	
		Quantitative analysis requires that we observe and monitor trends in features, such Raman modes, within and across a range of spectra. However this can be prohibitive when each spectrum contains several thousand data points. Principal component analysis (PCA) is a widespread technique in spectroscopy and chemometrics to reduce the dimensionality and facilitating the quantification of such information\cite{Jolliffe2016}. PCA forms a vector space for the data set whose basis vectors are themselves mutually orthogonal, linear functions of the original data, and which maximize variability within the set. 
	
		The basis vectors are functions of the data set and are excellent tools to reveal trends which may not be apparent upon manual inspection. These trends can be put to use in predictive multivariate linear regression models like principal component regression (PCR), significantly simplifying and expediting quantitative analysis of Raman spectra. 
	
		However, there are limitations on PCA whose basis vectors are highly sensitive to outliers in the data set and may unintentionally correlate with experimental artifacts such as a fluorescent background, detector noise, and drift-based amplitude variations. The PCR's capability to repeatedly provide reliable and meaningful predictions can be significantly hampered by the presence of these spectral features. Such features are omnipresent in Raman spectra obtained from real-life biofluids. Correcting and accounting for artifacts is of particular importance in biological analysis where samples are inherently variable from one patient to the next \cite{Jarvis2005}. Compensating for these aberrations with a pre-processing step can drastically improve PCR's ability to form a robust model. PCR also requires a decision about how many basis vectors to include in the model. If too few are included, trends in the data set may not be sufficiently characterized. Higher order basis vectors often describe noise, so these should be excluded. The number of vectors must be optimized to minimize computation time and artifact correlation while maximizing prediction accuracy. 
	
		In order to mitigate some of these limitations, cross-validation is a standard technique to predict PCR model performance by assessing the response of a PCR model on different spectra using varying combinations of basis vectors. Selecting the number of vectors may be based on distributional assumptions about the training set or by measuring the degree of fit. However, these methods can fail if the distribution is incorrect and may not describe the model's performance for new spectra \cite{Diana2002} \cite{Hoskuldsson2001}. The possible combinations of loading vectors and pre-processing methods for cross-validation rapidly becomes prohibitive to assess, particularly for large data sets. The optimal pre-processing method for a training set can be determined by trial and error, but this is not a generic or a robust approach and does not indicate the suitability to future predictions. It also proves to be a lengthy process when multiple techniques are being tested simultaneously \cite{Heraud2006} \cite{Gerretzen2015}. As such, there is still a real lack of a dependable, automated and sufficiently generic approach to render PCR immune from its limitations and enable it to fulfill its power of quantitatively analyzing Raman spectra, from real life biofluidic samples. 
	
		In this work we have developed a two step process to quickly determine the suitability of pre-processing methods and output the optimal number of basis vectors for a robust predictive model. For a given pre-processing method we perform cross-validation on a known data set, and analyze the predicted residual matrix using a statistical F-test. F-tests have previously been used in biological context to determine the number of significant factors in a model, but not to determine pre-processing suitability \cite{Malinowski1989} \cite{Diana2002} \cite{Keithley2010}. Our approach is validated using the analysis of two fluids designed to mimic human tears and whole human blood. The combination of LCWs and a statistical F-test to optimize the PCR model allows us to reliably predict constituent concentrations in tertiary biological phantom solutions with a sensitivity exceeding 0.1 mg/mL, bringing the predictions into a diagnostically-relevant range. To the best of our knowledge this is the first time statistical F-testing is being used to determine the suitability of a pre-processing technique and simultaneously output the optimal number of basis vectors for a predictive model.

	\section{\label{sec:formalismPCR}Formalism of Principal Component Regression}
		
		the relationship between PCs and the original data set is best illustrated by inspecting Equation \ref{eq:pcr_maineq}. Each spectrum in the set can be expressed as a linear sum of $k$ principal components, which are themselves the product of a basis or loading vector and an associated weighting factor. 
	
		\begin{equation}
		\begin{split}
			\label{eq:pcr_maineq}
			\mathbf{X} 	&= \mathbf{TP}^\intercal + \mathbf{E} \\
			&= \mathbf{t}_1\mathbf{p}_1^\intercal + ... + \mathbf{t}_k\mathbf{p}_k^\intercal + \mathbf{E} \\		
			\end{split}
			\end{equation}
			\begin{equation*}
			\begin{split}
			\mathbf{P}	&: \text{Loading vectors ($j \times k$)}\\
			\mathbf{T}	&: \text{Scores ($i \times k$)}\\
			\mathbf{X}	&: \text{Spectra ($i \times j$)}\\
		\end{split}
		\end{equation*}
	
		Expressing a data set with respect to its principal components presents us with the opportunity to form a regression prediction model used to gain insight into another spectra's unknown composition. Throughout this discussion we will use the convention of upper-case bold to denote matrices, lower-case bold for column vectors, and lower-case italics for scalars. A superscript ($^\intercal$) is used to indicate the matrix or vector transpose. In this study the predicted items in PCR are concentrations of an analyte, however the process is equally applicable to a wide range of predictions such as solubility of peanut proteins, or positioning error in GPS \cite{Wang2017} \cite{Adusumilli2015}.
	
		Two independent data sets are required for PCR: a reference set whose concentrations are known, and a set whose concentrations are unknown. Starting with the data set with known composition, the algorithm NIPALS (non-iterative partial least squares) is used to obtain the weighting and loading vectors. Score, loading, and concentration matrices are used as shown in equations \ref{eq:pcr_regr} and \ref{eq:pcr_concEst} to obtain the regression coefficients and an estimation for constituent concentrations ($\mathbf{C_{est}}$) \cite{Geladi1986}.
	
		\begin{equation}
		\label{eq:pcr_regr}
			\mathbf{B} = \frac{\mathbf{T}^\intercal \mathbf{C}}{\mathbf{T}^\intercal \mathbf{T}}
		\end{equation}
		\begin{equation}
		\label{eq:pcr_concEst}
		\begin{split}
			\mathbf{C_{est}} 	&= \mathbf{B} \mathbf{T}^\intercal \\
			&= \mathbf{B} (\mathbf{X_{est}}\mathbf{P})^\intercal
		\end{split}
		\end{equation}
		\begin{equation*}
		\begin{split}
			\mathbf{B}	&: \text{Regression coefficients ($q \times k$)}\\
			\mathbf{C}	&: \text{Known concentrations ($q \times i$)}\\
			\mathbf{C_{est}}	&: \text{Estimated concentrations ($q \times r$)}\\
			\mathbf{X_{est}}	&: \text{Spectra to be estimated ($r \times j$)}\\
		\end{split}
		\end{equation*}
		
		$\mathbf{X_{est}}$ consists of $r$ rows of independent spectra to be estimated. \textbf{C} contains $q$ unique species whose concentrations are to be calculated for $\mathbf{C_{est}}$. 
	
	\section{\label{sec:mat and methods}Materials and Methods}	
	
		Here we shall describe the techniques of pre-processing, principal component determination, and cross-validation. These methods on their own are commonplace in the processing of Raman spectra. However the decision of which technique is most suitable for pre-processing requires considerable manual intervention, and thus the model's robustness will vary with the operator. Our method seeks to rectify this. We have used the pre-processing techniques described in section \ref{sec:preproc} in conjunction with statistical analysis methods in section \ref{sec:stats Ftest} to develop a system which can optimize the pre-processing method without user input. With these adaptations multiple users with varied experience levels can operate a Raman analysis system with the confidence that their results hold significance. Many biofluids can be processed under a single system without requiring manual calibration for each set of results. Raman spectra of a biofluid collected on multiple systems can be used together for one common diagnostic analysis, regardless of spectral variations caused small changes in measurement conditions. We will demonstrate in this study that the technique can repeatedly generate a robust predictive model, opening up exciting new opportunities for the adoption of Raman spectroscopy in biological applications. 
	
	\subsection{\label{sec:preproc}Preprocessing Techniques}
		\subsubsection*{Intensity Normalization}
			Amplitude normalization is often required prior to PCR to ensure consistency between spectra. Between experiments, or even within one, slight changes in path-length or focal point can manifest as differences in the amplitude of Raman peaks, negatively impacting the results of PCR. Three commonly-used techniques to accomplish amplitude normalization are scaling based on reference peak height, by standard normal variate transform (SNV) (equation \ref{eq:pca_snv}), or with the robust normal variate transform (RNV) (equation \ref{eq:pca_rnv}) \cite{Zeaiter2005}.
			
			\begin{equation}
			\label{eq:pca_snv}
				\mathbf{x_{adj}} = \frac{\mathbf{x}-\mu(\mathbf{x})}{\sigma(\mathbf{x})} 
			\end{equation}
			
			SNV centers each spectrum in a set by its mean ($\mu$) and scales it by its standard deviation ($\sigma$). RNV centers and scales spectral data similarly, but makes an additional assumption about its distribution. RNV was introduced as the solution to a closure problem. Closure, in statistics, indicates that the sum of the set is equal to a certain number \cite{Guo1999}. Therefore if one value in the set increases, another must decrease to compensate. Each SNV spectrum is mean-centered therefore its sum is zero and the set is closed. This can have the unwanted effect of introducing artificial variance in an attempt to maintain closure \cite{Guo1999}. RNV modifies SNV by using percentiles (denoted $pct$) rather than means:
			
			\begin{equation}
			\label{eq:pca_rnv}
				\mathbf{x_{adj}} = \frac{\mathbf{x} - pct(\mathbf{x})}{\sigma[\mathbf{x} \leq pct(\mathbf{x})]} 
			\end{equation}
			
			An immediately apparent benefit to RNV is reduced influence of outliers. If the distribution of response measurements for one variable is skewed in any way, the percentile is a more robust descriptor of tendencies within the data set than the standard deviation \cite{mcclave2012statistics}. The percentile must be optimized in order for RNV to provide the best possible prediction model; several techniques have been proposed to accomplish this and can be found in literature \cite{Guo1999}.
			
		\subsubsection*{Smoothing}
			Noise in the Raman signal can greatly affect the robustness of of a PCR model. We desire a model which describes the Raman response, not the artifacts within it. Savitzky-Golay (SG) is the best-known smoothing algorithm \cite{Zeaiter2005}. This method uses a moving window of points which are successively fitted to a polynomial using the method of least squares. Each point in the data set is evaluated and replaced with the value of the polynomial which is the best fit based on the group of points evaluated \cite{Savitzky1964}. The SG algorithm is well-suited to removing noise, but can affect the spectrum in some points where sharp peaks are present. The size of the moving-window and degree of the polynomial must therefore be optimized on a case-to-case basis.
			
		\subsubsection*{Differentiation}
			Systematic trends in a spectrum may be compensated for with differentiation. A constant offset baseline is removed with the first derivative, linearly sloped baseline via the second derivative, and so on \cite{Zeaiter2005}. 
	
	\subsection{\label{sec:NIPALS}Calculation of Principal Components}
	
		If all PCs are desired, singular value decomposition is a computationally-efficient option. When only the PCs of highest variance are desired, NIPALS has been shown to reduce computation time \cite{Wold1987}. NIPALS iteratively calculates each set of weights and loadings by repeatedly regressing $\mathbf{X}$ on to a weighting matrix to obtain improved loading vectors, and then regressing the $\mathbf{X}$ on to the improved loading vectors to obtain improved weightings. Once the algorithm converges, the calculated PC is subtracted from $\mathbf{X}$ and the residual data matrix is used to calculate successive components \cite{Andrecut2009}. NIPALS may be terminated once sufficient variance is described \cite{Geladi1986}.
		
	\subsection{\label{sec:stats Ftest}Cross-Validation and Statistical F-Testing}
				
		Higher order PCs can represent noise or experimental artifacts within the data set, and may not pose a positive contribution to the predictive model. Including these variables introduces risks such as over-fitting to the parameter being predicted, or compromising the model's performance when applied to new samples \cite{Andersen2010}. We have chosen leave-one-out cross validation to assess the predictive ability of PCR models. It is an excellent tool to account for sample-specific errors and generate a reproducible predictive model \cite{Bro2008}. Though it can be computationally expensive in large data sets we shall demonstrate how the addition of a statistical F-test can pre-emptively reject pre-treatment methods which lead to an ineffective model. 
		
		Leave-one-out cross-validation is initiated with the matrix $\mathbf{X}$, calibration spectra whose concentrations are known. In turn each spectra is removed and the remaining $i-1$ spectra are used to predict the composition of removed spectra using $1$ through $i-2$ PCs. Exceeding $i-2$ principal components presents the possibility that each spectra forms its own basis vector, negating PCA dimension reduction. Following PCR, error is calculated using Predicted Residual Sum of Squares (PRESS): 
		
		\begin{equation}
		\label{eq:pca_press}
			PRESS = \sum_{n=1}^{q} (\mathbf{C_{est}}_{,n} - \mathbf{C}_n)^2
		\end{equation}
		
		Full leave-one-out cross-validation yields a matrix $\mathbf{S}$ sized $i \times i-2$ of PRESS values on which we perform a statistical F-test to compare variability within one PC to that between different PCs \cite{mcclave2012statistics}. An F-test gives a statistical measure of whether differences in prediction are significant or simply due to sampling. This is essential during cross-validation in order to assess the model's suitability. 
		
		Significance determination begins with the null hypothesis ($H_0$) that the mean PRESS values for each PC are equal and an alternative hypothesis ($H_a$) that they are not all equal. A measure of the variability within each PC is first calculated using a modified sum of squares for error (SSE) (equation \ref{eq:pca_SSE}). A measurement for the variation within $\mathbf{S}$ caused by using different PCs is explained using the sum of squares for treatments (SST) (equation \ref{eq:pca_SST}).
		
		\begin{equation}
		\label{eq:pca_SSE}
			SSE = \sum_{n=1}^{i-1} \sum_{m=1}^{i-2} (\mathbf{S}_{n,m} - \overline{\mathbf{S}}_m)^2
		\end{equation}
		\begin{equation}
		\label{eq:pca_SST}
			SST = (i-2) \sum_{m=1}^{i-2} (\overline{\mathbf{S}}_m - \overline{\mathbf{S}})^2
		\end{equation}
		
		The F-statistic (equation \ref{eq:pca_f-ratio}) is determined from SST and SSE, after normalizing each factor by their respective degrees of freedom. 
		
		\begin{equation}
		\label{eq:pca_f-ratio}
			F = \frac{SST/(i-3)}{SSE} 
		\end{equation}
		
		Statistical significance of $F$ is determined by referencing a table of values for a chosen significance level $\alpha$, the probability of erroneously rejecting the null hypothesis. If $F > F_{\alpha}$ we reject $H_0$. To perform F-test computations using equations \ref{eq:pca_SSE}-\ref{eq:pca_f-ratio} and determine significance we are using MatLAB's \texttt{anova1} function from the Statistics and Machine Learning Toolbox. 
	
\begin{figure}
			\centering
			\includegraphics[width=3in]{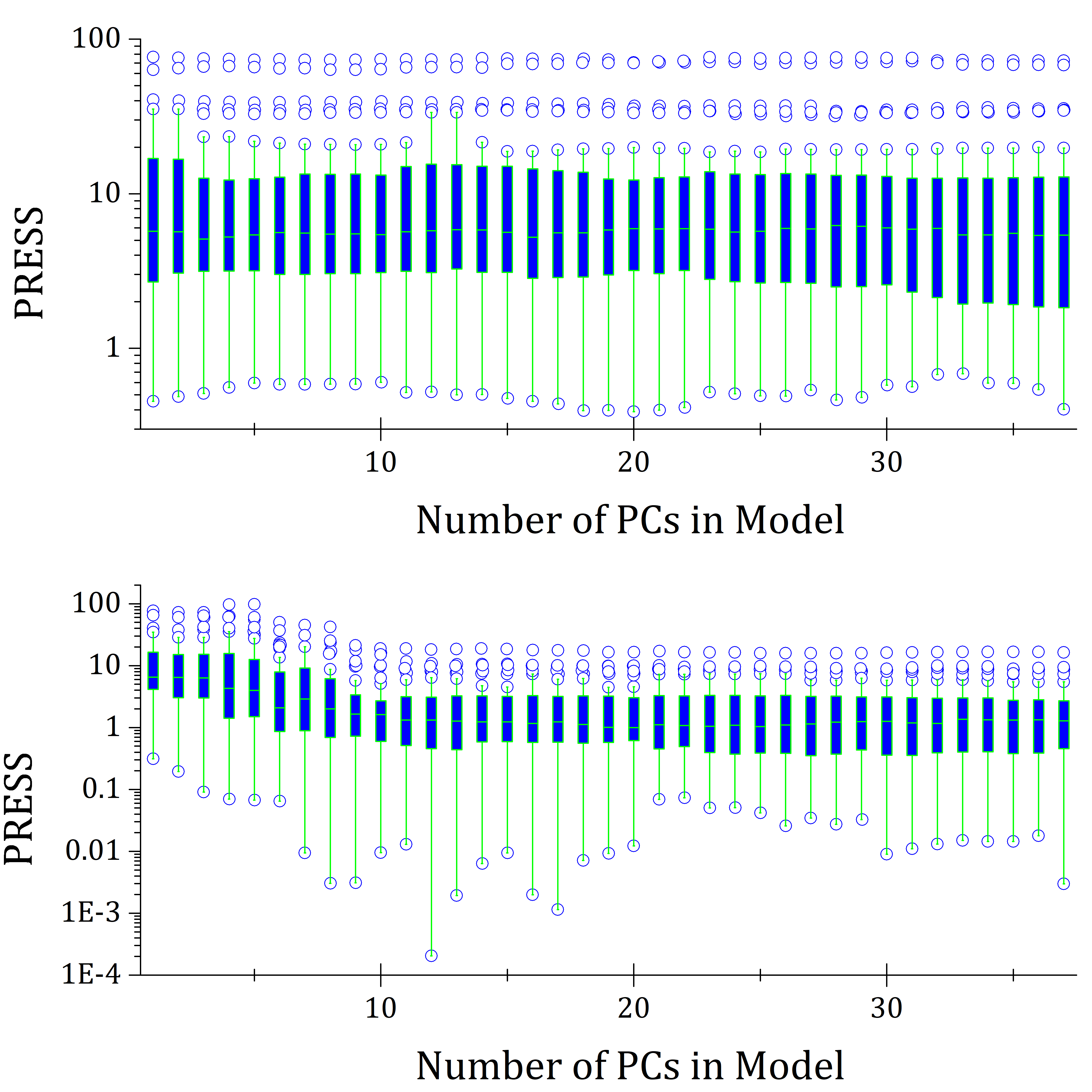}
			\caption{\label{fig:press Sig Both} Variation within PRESS values for each number of PCs, caused by using a different Raman spectra as the unknown in a cross-validation training model. Top: Spectra pre-processed using the first derivative. Little change is observed in PRESS values as the number of PCs is altered, indicating that the pre-processing method is ill-suited for a PCR model. Bottom: Preprocessing using baseline removal and RNV at the $90^{\text{th}}$ percentile. As the number of PCs is altered, PRESS values consistently improve significantly. We expect that this pre-treatment method will yield a well-performing PCR model.}
		\end{figure}

		When a different number of PCs leads to a statistically significant difference in concentration prediction ($F > F_{\alpha}$), we single out the PCs which show significance relative to the PC with highest prediction error. From this subset we select the PC with combined lowest error in prediction and highest measure of significance. In the case that we fail to reject $H_0$ the user is alerted and we opt for the PC with lowest sum of PRESS values. We have observed that failure to reject $H_0$ is not a function of the data set itself. Instead it indicates an inappropriate selection in pre-processing treatment. Thus this technique allows us to immediately reject a pre-processing method and rapidly proceed to the processing techniques most suited to the particular data set. Figure \ref{fig:press Sig Both} illustrates this point. Both plots begin with an identical data set. The data in the top plot is pre-processed using first derivative and the data in the bottom plot using baseline removal and RNV at the $90^{\text{th}}$ percentile. A difference in pre-treatment method leads the second plot to significantly better predictions as a function of the number of PCs which is mirrored in the resulting prediction accuracy.
		
		Variability in the PRESS matrix will occur in two dimensions: for each file as the unknown in the cross validation model there will be variation as a function of the number of PCs, and within each PC thee will be variation as a function of the file which is used as the unknown. We demonstrate this variability using a box plot. A filled rectangle shows the interquartile range (IQR), the area between 25th and 75th percentiles, with the median value marked within. A thin solid line extends to the minimum and maximum points which are not considered outliers, which are depicted as open circles and defined by the criteria that they lie further from the 25th or 75 percentile than 1.5 times the IQR \cite{mcclave2012statistics} All F-tests are evaluated at significance level $\alpha = 0.05$.
		
	\subsection{\label{sec:RamanCollection}Spectra Collection}
		
		All Raman spectra were collected on an Horiba HR800 Spectrometer using a grating of 600 grooves/$mm$ and an excitation laser centered at 633 $nm$. Laser power was 17 mW. Exposure times ranged from 60$s$ to 240$s$ with 1-3 acquisitions depending on desired SNR. Fluidic sample collection was assisted and enhanced by a custom-built microfluidic device with integrated Teflon capillary tube LCW. Details on liquid-core waveguide enhancement of fluidic samples can be found in our previous research \cite{Eftekhari2011} \cite{Mak2013}.  
	
		\begin{figure}
			\centering
			\includegraphics[width=3in]{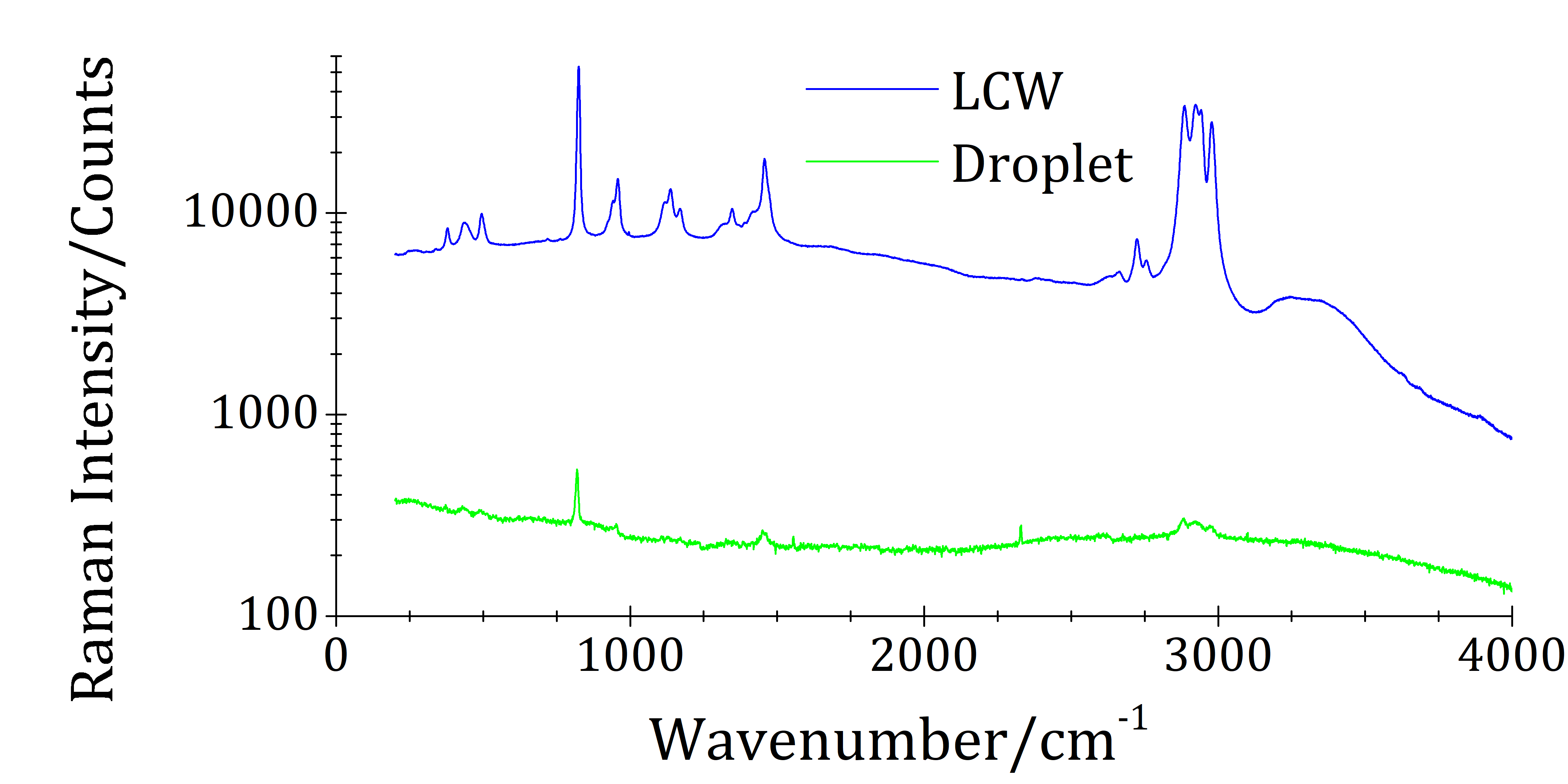}
			\caption{\label{fig:RamanMethodCompare}Comparison of Raman spectra of isopropyl alcohol collected with two different methods. The LCW provides substantial enhancement under otherwise identical collection conditions.}
		\end{figure}
		
\begin{figure}
			\centering
			\includegraphics[width=3in]{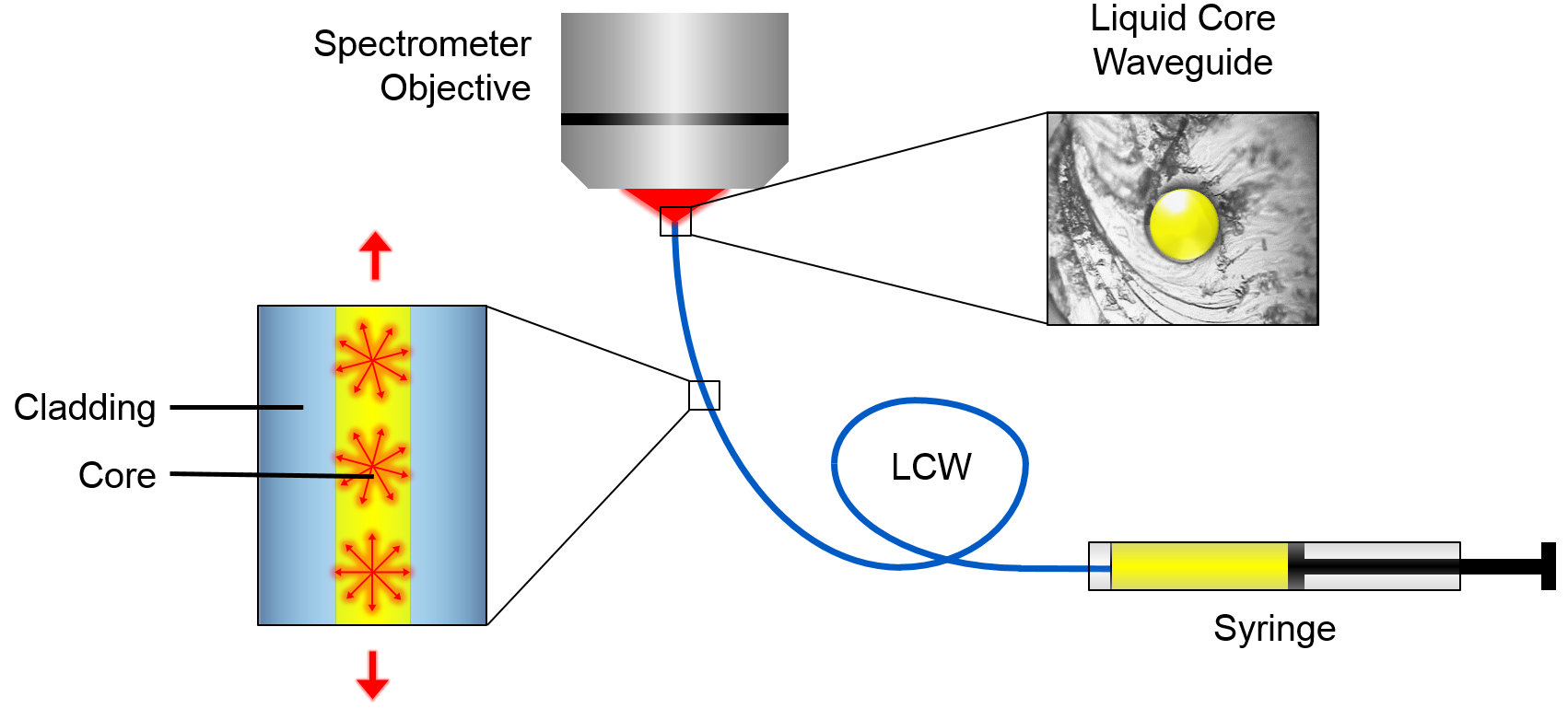}
			\caption{\label{fig:RamanLCW}LCW Spectrometer configuration}
		\end{figure}

		Figure \ref{fig:RamanMethodCompare} illustrates the enhancement that a LCW can provide over conventional Raman collection using a drop of sample. The microfluidic device eliminates evaporation, stabilizing the focal point and enabling us to be confident that the biofluid composition remains unchanged for the duration of lengthy collection times. Our microfluidic LCW technique, shown in figure \ref{fig:RamanLCW}, has allowed us to obtain a collection of high quality Raman spectra to demonstrate our statistical technique and its applicability to physiologically relevant biofluid analysis. Further reading on LCWs can be found in previous publications \cite{Eftekhari2011} \cite{Mak2013}.
		
	\subsection{Biofluids}
		
		Raman spectra of biological samples are inherently variable not only between patients, but within an individual due factors such as collection method or hydration state \cite{Sen1980} \cite{Chiappin2007} \cite{Berger2010}. Abnormal concentrations of certain analytes are diagnostically relevant, so it is crucial that we are able to compensate for these variations with a high degree of reproducibility in order to accurately assess one patient relative to a population. Two artificial biofluid phantoms were constructed for our experiments which satisfied either diagnostic relevance or optical property matching, and which eliminated special training, handling, and disposal requirements.
		
		Human tears contain upwards of 400 unique proteins \cite{Filik2009}. Variations in the concentration of the protein lysozyme in particular have been shown to be relevant for the diagnosis of herpes simplex virus, Sjoren's syndrome, and kerato-conjunctivitis \cite{Eylan1977} \cite{Janssen1983}. In addition to proteins, tears contain diagnostically relevant substances such as glucose. Diabetes is one of the most prevalent diseases among developed nations, affecting hundreds of millions worldwide \cite{Bispo2013}. It requires regular blood glucose monitoring, and facilitating this process has been the source of extensive research \cite{Qi2007}. Glucose is known to be difficult to detect in Raman spectroscopy due to its small scattering cross-section \cite{Yang2012}. We demonstrate in this experiment that our liquid core waveguide microfluidic device, when used together with this statistical analysis method, allows us to repeatedly and accurately predict physiologically relevant concentrations of glucose and lysozyme in human tears.
		
		Samples to mimic human tears were comprised of lysozyme and glucose dissolved in deionized water (DIW), each at physiologically relevant levels. This range for lysozyme is between 0 and 10 mg/ml, and for glucose it is between 0 and 1 mg/mL \cite{Eylan1977} \cite{Sen1980} \cite{Filik2009}. D-(+)-Glucose (G8270) and Lysozyme from chicken egg white (62970) were purchased from Sigma-Aldrich. DIW was obtained from a Milli-Q in the Bahen Prototyping facility at Toronto Nanofabrication Center (TNFC). 
	
		To simplify blood analysis we have focused on the optical properties of whole blood. Diagnostics using whole blood minimizes the dependence on specialized equipment to pre-process the sample, but detecting low-concentration analytes is severely compromised due to red blood cell's strong scattering coefficient \cite{Enejder2002}. 
	
		As a substitute for whole-blood we have used a 20\% Intralipid fat emulsion (I141) purchased from Sigma-Aldrich. The scattering coefficient of this media is a good match to that of whole blood \cite{Bosschaart2014} \cite{Michels2008}. Additionally, intralipid is inexpensive, non-toxic, and requires no special handling or disposal, making it a suitable phantom for this proof-of-concept study. Two solutes were used within intralipid: DIW and isopropyl alcohol (IPA). DIW was obtained from TNFC, and IPA (190764) from Sigma-Aldrich. These solutes were selected due to their availability, in order to demonstrate that our liquid core waveguide microfluidic device allows us to detect increasingly lower concentrations within an opaque solution, and the statistical analysis method reliably distinguishes different Raman modes when more than one analyte is present. 
		
		\begin{figure}
			\centering
			\includegraphics[width=3in]{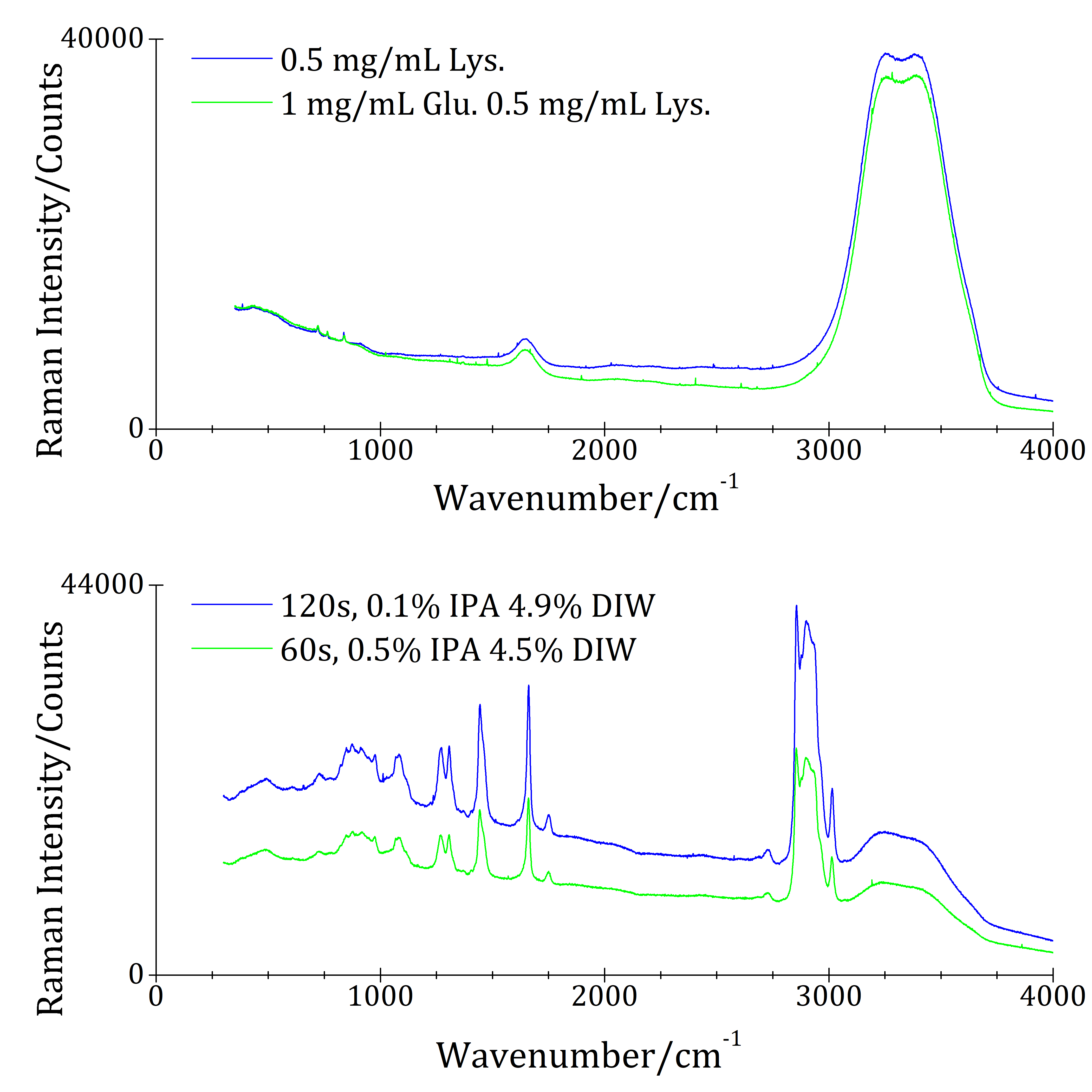}
			\caption{\label{fig:intralipidRamanVariation}Top: Two spectra of artifical tears, with composition varying by 1 mg/mL glucose and 0.5 mg/mL lysozyme. Baseline drift requires correction. Bottom: Two spectra of intralipid, with composition varying by 0.4\% IPA and 0.4\% DIW, collected with different exposure times. Amplitudes require normalization.}
		\end{figure}
		
		Enhanced Raman spectra of biofluids are an ideal candidate to demonstrate our improved PCR robustness algorithm. Fluorescent baseline (figure \ref{fig:intralipidRamanVariation} top) and amplitude variations (figure \ref{fig:intralipidRamanVariation} bottom ) are omnipresent in the Raman spectra of biofluids. Including them in a PCR model in raw form may worsen predictions by correlating insignificant amplitude variations, despite their potential to significantly strengthen the model if artefacts can be corrected. Using the methods described in section \ref{sec:mat and methods} our method successfully combats these key challenges which may prevent Raman spectroscopy from being considered as a candidate for reliable diagnostics.  
		
		\begin{figure}
			\centering
			\includegraphics[width=3.5in]{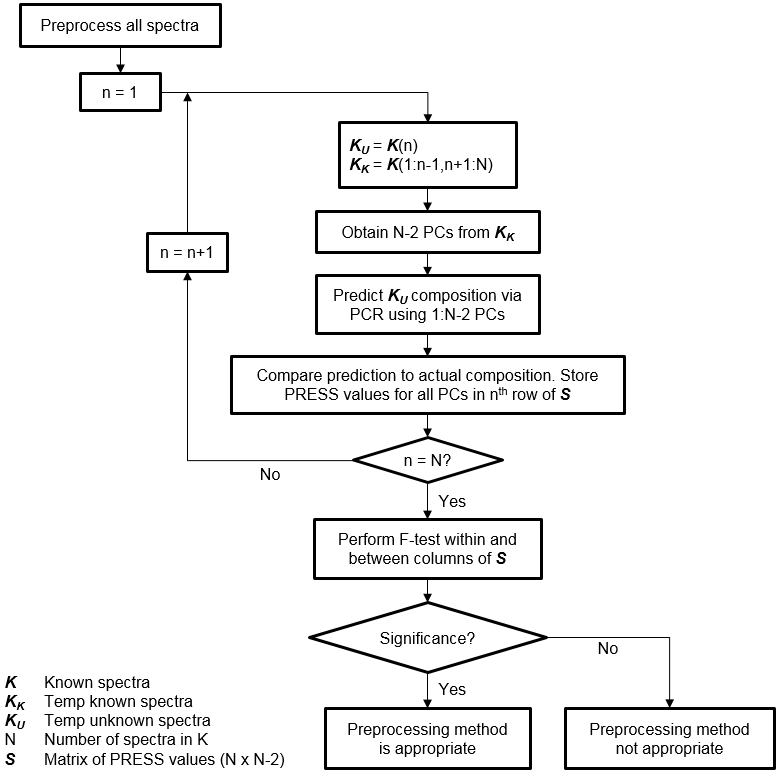}
			\caption{\label{fig:processFlow}The flow chart of pre-processing statistical significance algorithm}
		\end{figure}
		
	\section {Results and Discussion}
		
		We aim to demonstrate that our technique is capable of reliably quantifying analytes in artificial tears and intralipid solutions. We shall compare PCR model performance in the case where pre-treatment produces statistical significance in cross-validation, to the case where a pre-processing method is not found to produce significance. Within each set of Raman specta one spectum acted as the unknown and cross-validation was performed on the remainder, in order to predict the suitability of different pre-processing methods. To demonstrate the algorithm's efficacy we plot the PRESS values compared to the actual residual sum of squares (RSS) of the unknown spectra. In each case analyzed we shall demonstrate that our technique is superior at identifying a suitable a pre-treatment method and number of PCs for an optimal PCR model. 
		
		\begin{figure}
			\centering
			\includegraphics[width=3in]{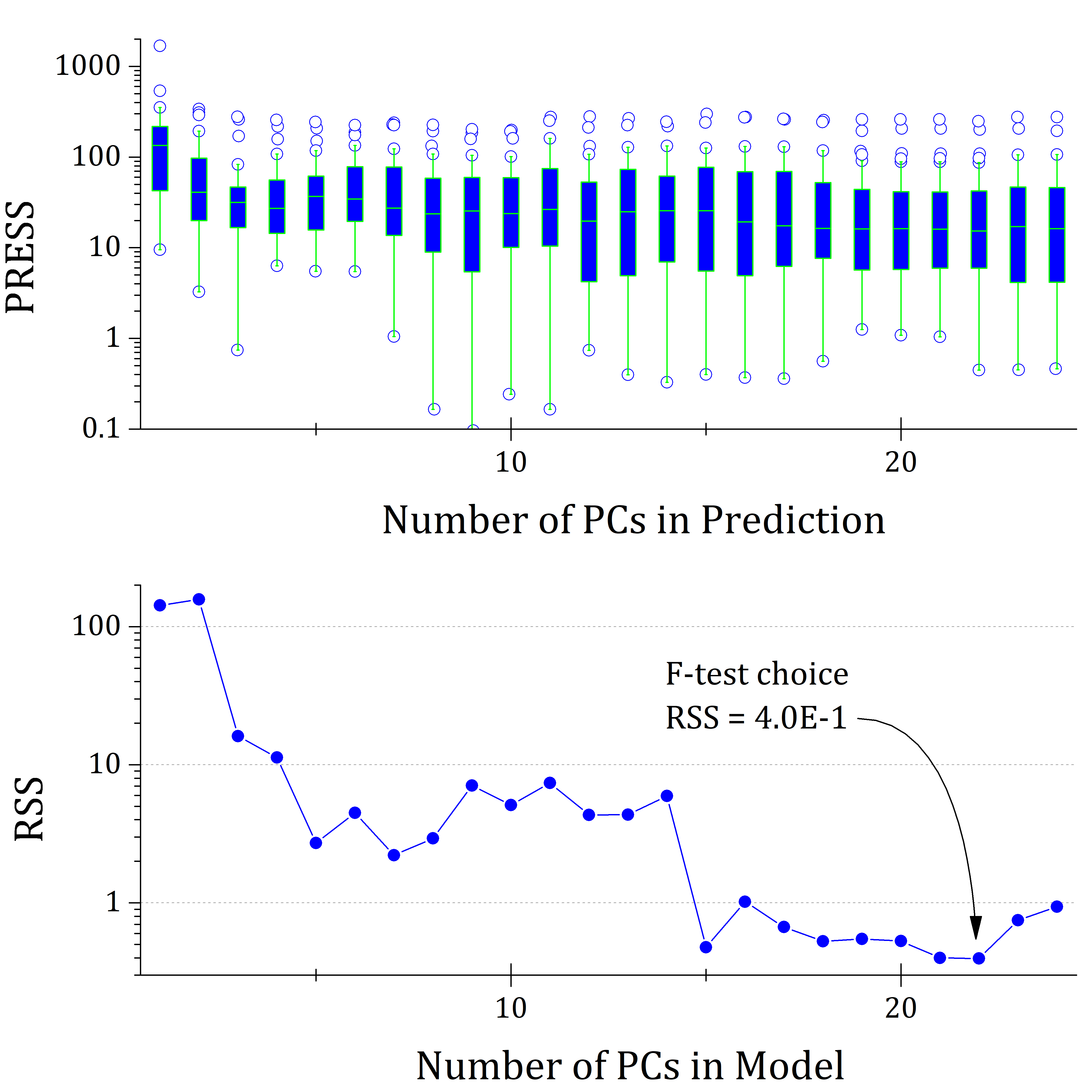}
			\caption{\label{fig:set1Significant}Intralipid solution, pre-processed with RNV (percentile 75) and baseline removal. PRESS values of cross-validation using 26 spectra (top) display significant variation with the number of PCs, and identifies 22 PCs for the regression model which is a minima in observed RSS (bottom).}
		\end{figure}
		
		Figure \ref{fig:set1Significant} depicts a case in which the pre-processing method shows statistical significance; cross validation model accurately predicts true RSS performance. The solutions from which these Raman spectrum were generated were designed to mimic optical properties of whole human blood and comprised intralipid, DIW and IPA. Preprocessing treatments applied were RNV at the 75th percentile after baseline removal. The data set contains 27 spectra, some of which were collected with longer exposure times, so amplitude normalization is necessary. Within the PRESS matrix each PC was cross-validated with 26 spectra. The box plot predicts an improvement of up to one order of magnitude of the median residual prediction. In addition, there are never more than four outliers for each PC during cross-validation. This data set passed the cross-validation test, and our F-test algorithm returned 22 as the optimal number of PCs for the data set based on the PRESS matrix significance. Actual composition for this solution is 5\% IPA, 85\% Intralipid, 10\% Deionized Water. Predicted concentrations with the model are 5.1\%, 84.6\%, 9.5\% respectively. The optimal PC value returned is indeed the minimum within actual RSS values, so we confirm the validity of the model formed with this pre-processing treatment. 
		
		\begin{figure}
			\centering
			\includegraphics[width=3in]{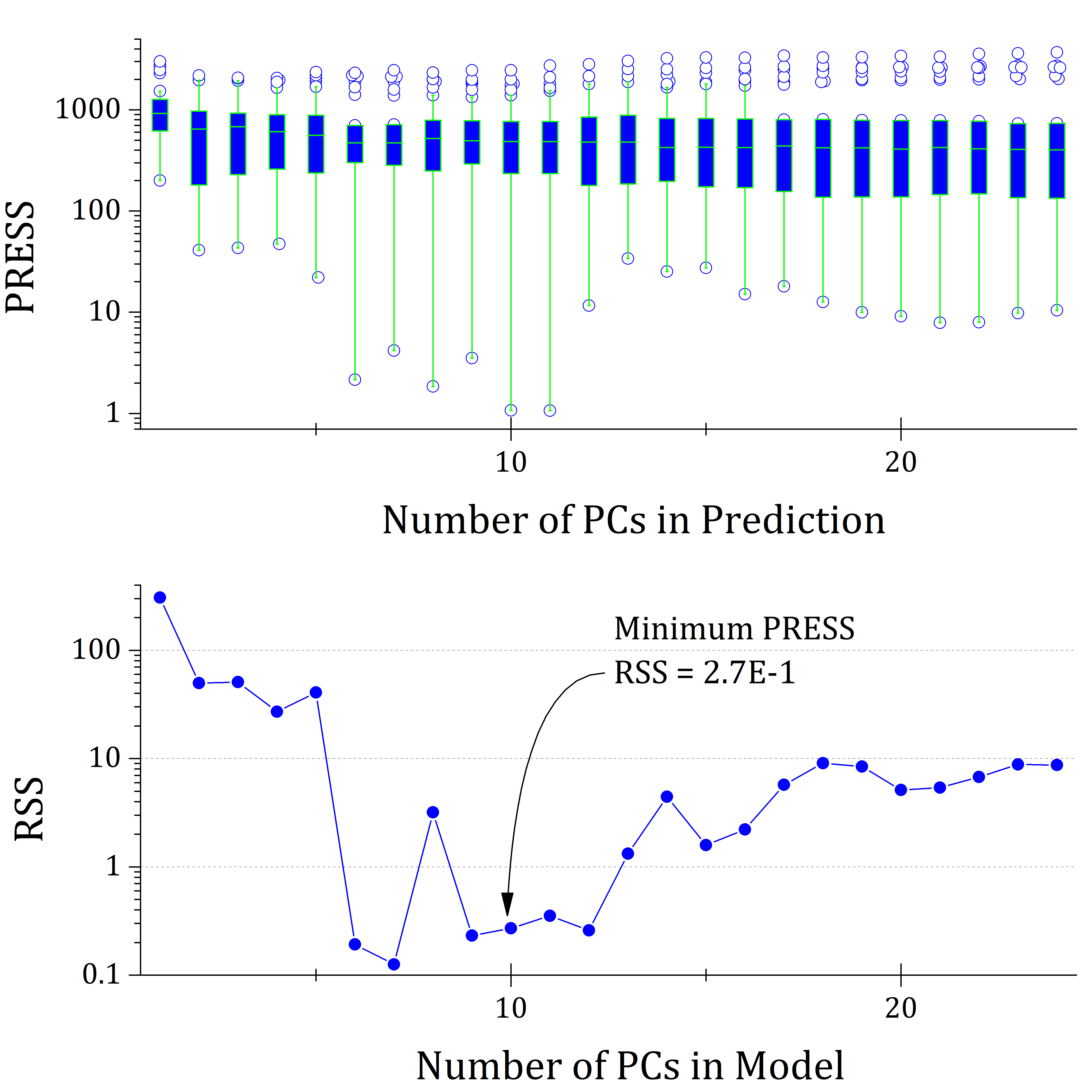}
			\caption{\label{fig:set1NotSignificant}Intralipid solution, pre-processed with SG smoothing and 2 derivatives. PRESS values of cross-validation using 26 spectra (top) do not significantly improve or worsen with the number of PCs. Choosing the PC with minimum sum of PRESS values does not lead to a model which minimizes observed RSS (bottom).}
		\end{figure}
		
		Figure \ref{fig:set1NotSignificant} depicts a case in which the pre-processing method fails the significance test; cross validation model fails to return true RSS performance. A similar data set from figure \ref{fig:set1Significant} was used for this case; a different Raman spectra as the unknown and cross-validation was performed with the remaining 26 spectra in the set. Preprocessing methods were SG smoothing and 2 derivatives. The box plot shows a greater number of outliers than figure \ref{fig:set1Significant}, and the median residual prediction improves by only one half order of magnitude as the number of PCs are varied. This pre-processing method did not pass the significance test. We select the PC with lowest sum of PRESS values (10) as the optimal number for the model, but using 6, 7, 9, or 12 PCs would have yielded lower RSS. Actual concentrations for this solution are 10\% IPA, 85\% Intralipid, 5\% Deionized Water. Predicted concentrations with the model are 10.3\%, 85.4\%, 5.3\%. The minimum sum of PRESS values is not a minima within actual RSS values; this pre-processing method's cross-validation model fails to identify four other PC values which would optimize the PCR model. 
		
		\begin{figure}
			\centering
			\includegraphics[width=3in]{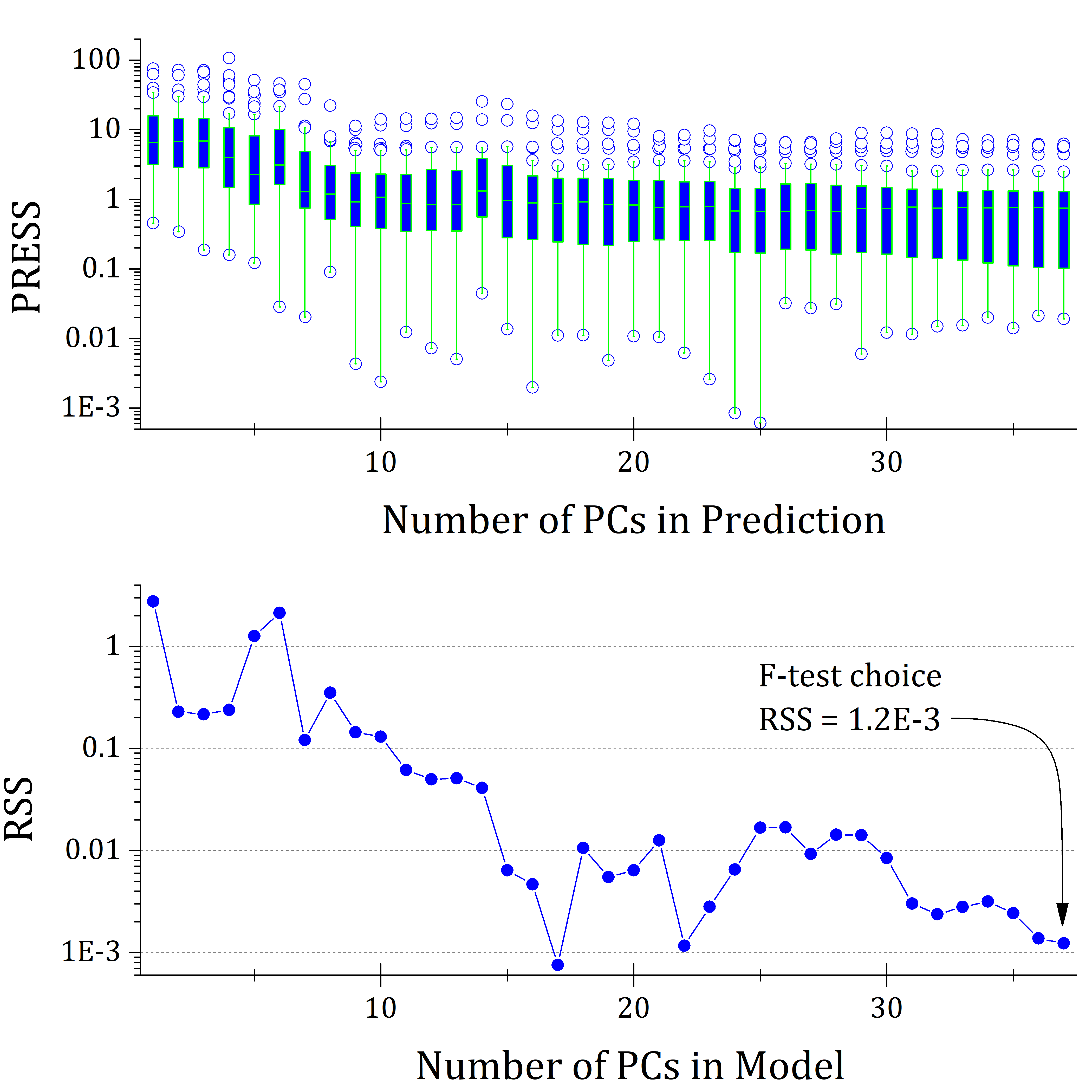}
			\caption{\label{fig:set2Significant}Artificial tear solution, pre-processed with baseline and cosmic spike removal. PRESS values of cross-validation using 39 spectra (top) significantly improve with the number of PCs, and the algorithm identifies 37 PCs for the regression model which is a minima in observed RSS (bottom).}
		\end{figure}
		
		A second case in which the pre-processing method shows statistical significance and cross validation predicts true RSS performance is illustrated in Figure \ref{fig:set2Significant}. 40 samples were designed to mimic human tears and consisted of DIW with varying concentrations of lysozyme and glucose, at physiologically relevant levels. All spectra have been pre-processed using a baseline and cosmic spike removal. Again we see more than one order of magnitude improvement in the median residual prediction. Up to 6 outliers are present in each PC, which is approximately the same proportion of the number of spectra in cross-validation that we observed in figure \ref{fig:set1Significant}. This data set passed the significance test and returned 37 as the optimal number of PCs. Actual composition is 0.5 mg/mL glucose and 0.5 mg/mL lysozyme. Predicted concentrations are 0.48 mg/mL and 0.52 mg/mL. The predicted optimal number of PCs is a local minima in actual RSS, validating the model and pre-processing method. 
		
		\begin{figure}
			\centering
			\includegraphics[width=3in]{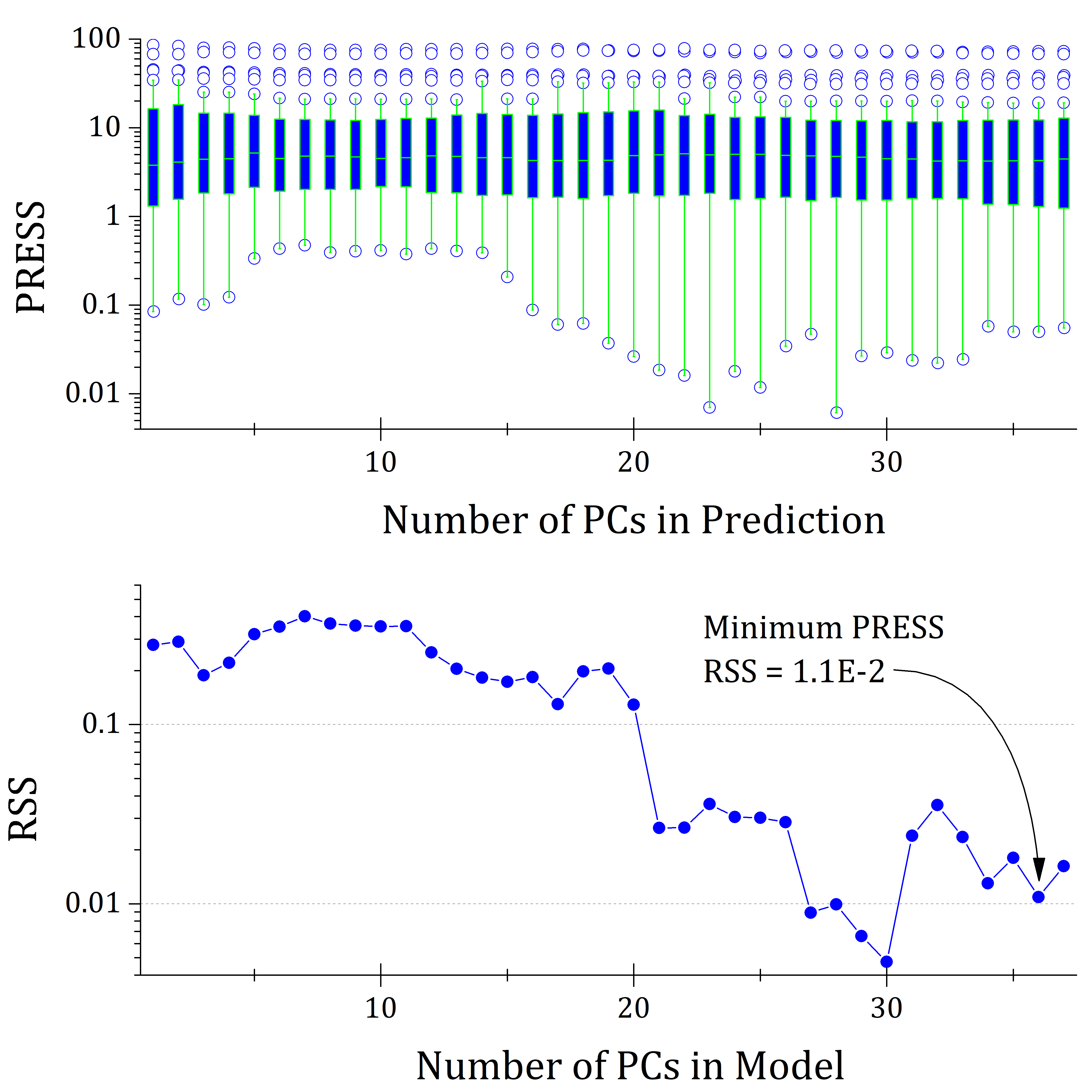}
			\caption{\label{fig:set2NotSignificant}Artificial tear solution, pre-processed with SG smoothing and 2 derivatives. PRESS values of cross-validation using 39 spectra (top) neither significantly improve nor worsen with the number of PCs. Choosing the PC with minimum sum of PRESS values does not minimize observed RSS (bottom).}
		\end{figure}
		
		Preprocessing fails to produce statistical significance in figure \ref{fig:set2NotSignificant}, leading to a predicted optimal PC which does not optimize the PCR model. A similar data set to the one in figure \ref{fig:set2Significant} is applied to this model using a different Raman spectrum as the unknown and the remaining 39 spectra in a cross-validation model. Preprocessing methods were SG smoothing and 2 derivatives. Residual prediction improves by less than 0.2 orders of magnitude as the number of PCs are varied. Again selecting the PC with lowest sum of PRESS values (36) as the optimal number for the model. Actual composition is 1 mg/mL glucose and 1 mg/mL lysozyme. Predicted concentrations are 0.94 mg/mL and 1.09 mg/mL. The minimum sum of PRESS does not return a minima in the RSS plot, confirming that this pre-processing technique is inappropriate for the data set. With our method we seek to contribute a tool which facilitates the most accurate PCR model available; we have shown in figures \ref{fig:set1NotSignificant} and \ref{fig:set2NotSignificant} that we can exclude these pre-processing treatments due to their PRESS matrices' lack of statistical significance, which manifests as a model that does not perform well when extended to new spectra.
		
\begin{figure}
			\centering
			\includegraphics[width=3in]{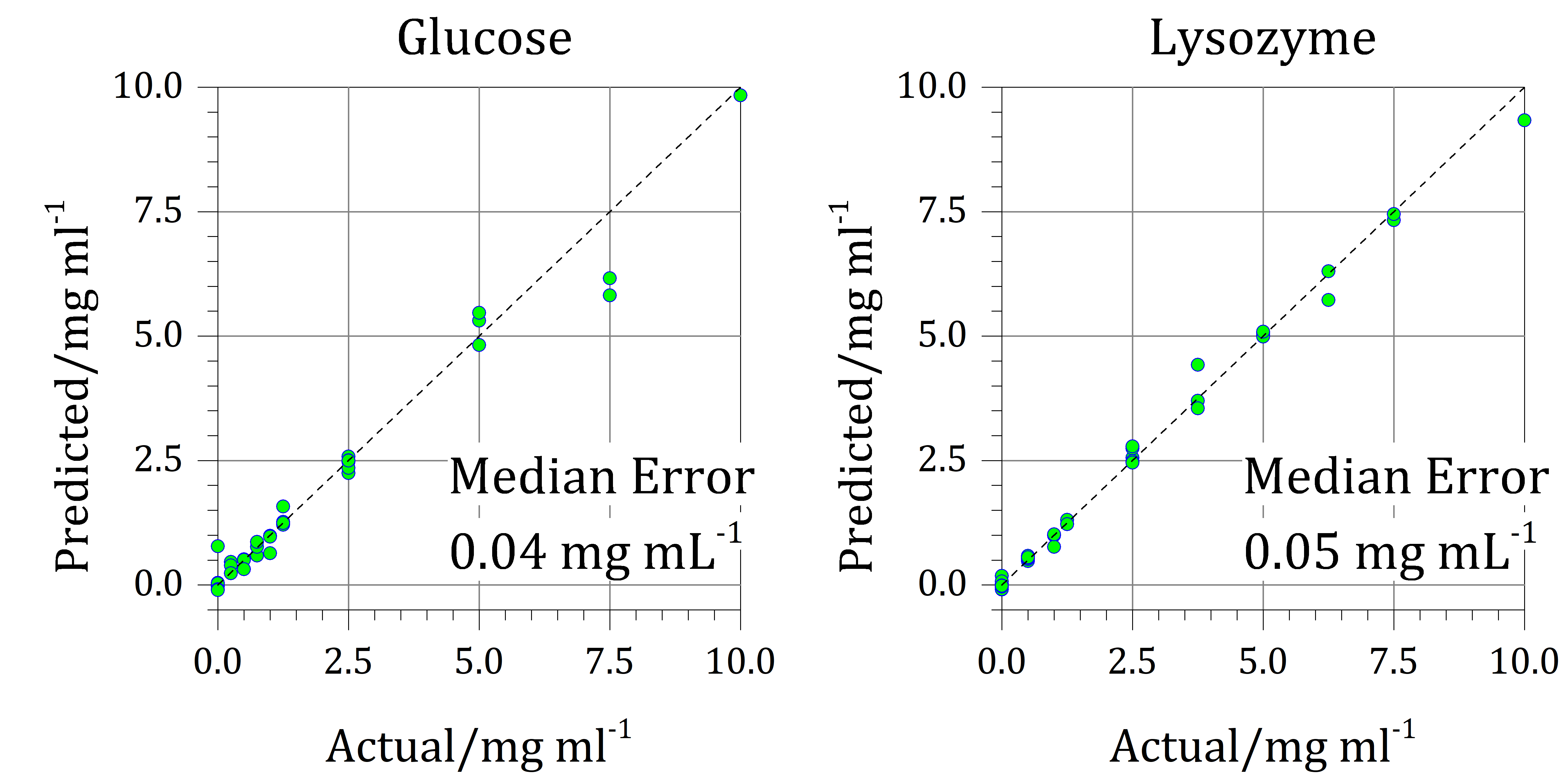}
			\caption{\label{fig:tearsBest}PCR prediction results compared to true concentration, for artificial tear solutions. Statistical significance is used to optimize the pre-processing method and number of PCs for each solution.}
		\end{figure}

		A general image of the performance provided using pre-processing choice influenced by statistical significance is shown in figure \ref{fig:tearsBest}. The pre-processing method for each unknown spectrum was optimized with the process outlined below:
		
		\begin{itemize}
			\item Determine the pre-processing techniques which produced statistical significance
			\item Identify the optimal number of PCs for each significant pre-processing method
			\item Select the pre-processing method with the lowest sum of PRESS values at the evaluated optimal number of PCs
		\end{itemize}
		
		Allowing our algorithm to select the pre-processing treatment for each individual unknown file allows us to obtain a median prediction error of 0.26 mg/mL for glucose, and 0.13 mg/mL for lysozyme. These error values fall either at or below the order of magnitude on which the solutes are physiologically relevant \cite{Eylan1977} \cite{Sen1980} \cite{Filik2009}. We are confident that further development and optimization of our LCW procedure and the statistical selection algorithm will allow us to bring these values well within the range required for reliable diagnostics.
		
		Differences in the statistical significance trend as a function of the pre-processing method in the previous examples serve to demonstrate the robustness of our analysis model. Existing techniques allow the operator to test and assess many pre-processing treatments, but the results do not provide any measure of the PCR model's performance when extended to future spectra. Statistical analysis using our method gives the user this extra information; current decisions about pre-processing method suitability lack the assurance factor that we can provide. 

	\section{Conclusion}
	
		Raman spectroscopy is an ideal tool for non-destructive analysis of biofluids, but insignificant variation between spectra must be compensated for in order to retrieve that which is diagnostically relevant. A multitude of pre-processing methods and combinations are available, and the optimal technique varies on a case-to-case basis depending on the analyte. 
		
		Our statistical F-test was applied to analyze variability within PRESS matrices of biofluid training spectra for PCR. This approach provides a confidence factor that, to the best of our knowledge, is not present in any other technique. Our algorithm is capable of determining the optimal pre-processing method, independent of the operator's experience level. The analysis technique has proven successful at identifying robustness of pre-processing techniques in two classes of artificial biofluids: a phantom of tears, and of whole human blood. Using this technique we demonstrate the ability to repeatedly detect and predict diagnostically-relevant concentrations of lysozyme in artificial tears. Within the same solutions we are within one order of magnitude of diagnostically-relevant glucose levels. Using an enhanced fluidic Raman technique and further algorithm optimization, we forecast that we will achieve and exceed the resolution necessary for glucose diagnostics in an upcoming experiment. 
		
		This method allows users of all skill levels to rapidly and confidently optimize pre-processing techniques for their data set, and dedicate time to diagnosis, detection, or data analysis techniques of biofluid Raman spectroscopy. We propose that this method successfully satisfies the need for a dependable, automated and sufficiently generic approach for PCR; limitations due to omnipresent biological variation are automatically compensated, allowing for a marked quality increase in robust quantitative analysis of Raman biofluid spectra.
		
	\section{Acknowledgements}
		
		This work was supported in part by Semiconductor Research Corporation's Global Research Collaboration (SRC-GRC) Program.

	\bibliography{statPaperArXiv}
	
\end{document}